\documentclass[aps,prl,preprint,tightenlines,superscriptaddress,showpacs,byrevtex]{revtex4}
\usepackage{graphicx} 
\usepackage{dcolumn}  
\usepackage{color}  

\usepackage{pstricks}
  
\newcommand{\ks}{K_S^0}
\newcommand{\ksks}{K_S^0K_S^0}
\newcommand{\aksks}{{\cal A}_{\ksks}}
\newcommand{\sksks}{{\cal S}_{\ksks}}
\newcommand{\sksksvalue}{-0.38}
\newcommand{\sksksstat}{0.77}
\newcommand{\skskssyst}{0.08}
\newcommand{\aksksvalue}{-0.38}
\newcommand{\aksksstat}{0.38}
\newcommand{\akskssyst}{0.05}

\newcommand{\kspizpiz}{K_S^0\pi^0\pi^0}
\newcommand{\akspizpiz}{{\cal A}_{\kspizpiz}}
\newcommand{\skspizpiz}{{\cal S}_{\kspizpiz}}
\newcommand{\skspizpizvalue}{+0.43}
\newcommand{\skspizpizstat}{0.49}
\newcommand{\skspizpizsyst}{0.09}
\newcommand{\akspizpizvalue}{-0.17}
\newcommand{\akspizpizstat}{0.24}
\newcommand{\akspizpizsyst}{0.06}

\newcommand{\qq}{q\overline{q}}
\newcommand{\de}{\Delta E}
\newcommand{\mbc}{M_{\rm bc}}
\newcommand{\dt}{\Delta t}
\newcommand{\dmd}{\Delta m_d}
\newcommand{\taub}{\tau_{B^0}}

\newcommand{\bbbar}{B^0\overline{B}{}^0}

\graphicspath{{ps}}

\begin{document}



\preprint{
  BELLE-CONF-0723
}

\title{ \quad\\[0.5cm]  
Measurements of $CP$ Violation Parameters
in $B^0\to\kspizpiz$ and $B^0\to\ksks$ Decays
}

\noaffiliation

\begin{abstract}
We present a measurement of the $CP$ violation parameters
in $B^0\to\kspizpiz$ and $B^0\to\ksks$ decays
using a data sample containing $657\times 10^6$ $B\overline{B}$
pairs collected with the Belle detector at the KEKB 
asymmetric-energy $e^+e^-$ collider operating at the $\Upsilon(4S)$
resonance.
We measure
\begin{eqnarray}
\skspizpiz &=& \skspizpizvalue \pm \skspizpizstat \pm \skspizpizsyst, \nonumber \\
\akspizpiz &=& \akspizpizvalue \pm \akspizpizstat \pm \akspizpizsyst, \nonumber \\
\sksks &=& \sksksvalue \pm \sksksstat \pm \skskssyst~~ {\rm and}, \nonumber \\
\aksks &=& \aksksvalue \pm \aksksstat \pm \akskssyst, \nonumber 
\end{eqnarray}
where the first and second errors are statistical and systematic,
respectively.
\end{abstract}


\affiliation{Budker Institute of Nuclear Physics, Novosibirsk}
\affiliation{Chiba University, Chiba}
\affiliation{University of Cincinnati, Cincinnati, Ohio 45221}
\affiliation{Department of Physics, Fu Jen Catholic University, Taipei}
\affiliation{Justus-Liebig-Universit\"at Gie\ss{}en, Gie\ss{}en}
\affiliation{The Graduate University for Advanced Studies, Hayama}
\affiliation{Gyeongsang National University, Chinju}
\affiliation{Hanyang University, Seoul}
\affiliation{University of Hawaii, Honolulu, Hawaii 96822}
\affiliation{High Energy Accelerator Research Organization (KEK), Tsukuba}
\affiliation{Hiroshima Institute of Technology, Hiroshima}
\affiliation{University of Illinois at Urbana-Champaign, Urbana, Illinois 61801}
\affiliation{Institute of High Energy Physics, Chinese Academy of Sciences, Beijing}
\affiliation{Institute of High Energy Physics, Vienna}
\affiliation{Institute of High Energy Physics, Protvino}
\affiliation{Institute for Theoretical and Experimental Physics, Moscow}
\affiliation{J. Stefan Institute, Ljubljana}
\affiliation{Kanagawa University, Yokohama}
\affiliation{Korea University, Seoul}
\affiliation{Kyoto University, Kyoto}
\affiliation{Kyungpook National University, Taegu}
\affiliation{Ecole Polyt\'ecnique F\'ed\'erale Lausanne, EPFL, Lausanne}
\affiliation{University of Ljubljana, Ljubljana}
\affiliation{University of Maribor, Maribor}
\affiliation{University of Melbourne, School of Physics, Victoria 3010}
\affiliation{Nagoya University, Nagoya}
\affiliation{Nara Women's University, Nara}
\affiliation{National Central University, Chung-li}
\affiliation{National United University, Miao Li}
\affiliation{Department of Physics, National Taiwan University, Taipei}
\affiliation{H. Niewodniczanski Institute of Nuclear Physics, Krakow}
\affiliation{Nippon Dental University, Niigata}
\affiliation{Niigata University, Niigata}
\affiliation{University of Nova Gorica, Nova Gorica}
\affiliation{Osaka City University, Osaka}
\affiliation{Osaka University, Osaka}
\affiliation{Panjab University, Chandigarh}
\affiliation{Peking University, Beijing}
\affiliation{University of Pittsburgh, Pittsburgh, Pennsylvania 15260}
\affiliation{Princeton University, Princeton, New Jersey 08544}
\affiliation{RIKEN BNL Research Center, Upton, New York 11973}
\affiliation{Saga University, Saga}
\affiliation{University of Science and Technology of China, Hefei}
\affiliation{Seoul National University, Seoul}
\affiliation{Shinshu University, Nagano}
\affiliation{Sungkyunkwan University, Suwon}
\affiliation{University of Sydney, Sydney, New South Wales}
\affiliation{Tata Institute of Fundamental Research, Mumbai}
\affiliation{Toho University, Funabashi}
\affiliation{Tohoku Gakuin University, Tagajo}
\affiliation{Tohoku University, Sendai}
\affiliation{Department of Physics, University of Tokyo, Tokyo}
\affiliation{Tokyo Institute of Technology, Tokyo}
\affiliation{Tokyo Metropolitan University, Tokyo}
\affiliation{Tokyo University of Agriculture and Technology, Tokyo}
\affiliation{Toyama National College of Maritime Technology, Toyama}
\affiliation{Virginia Polytechnic Institute and State University, Blacksburg, Virginia 24061}
\affiliation{Yonsei University, Seoul}
  \author{K.~Abe}\affiliation{High Energy Accelerator Research Organization (KEK), Tsukuba} 
  \author{I.~Adachi}\affiliation{High Energy Accelerator Research Organization (KEK), Tsukuba} 
  \author{H.~Aihara}\affiliation{Department of Physics, University of Tokyo, Tokyo} 
  \author{K.~Arinstein}\affiliation{Budker Institute of Nuclear Physics, Novosibirsk} 
  \author{T.~Aso}\affiliation{Toyama National College of Maritime Technology, Toyama} 
  \author{V.~Aulchenko}\affiliation{Budker Institute of Nuclear Physics, Novosibirsk} 
  \author{T.~Aushev}\affiliation{Ecole Polyt\'ecnique F\'ed\'erale Lausanne, EPFL, Lausanne}\affiliation{Institute for Theoretical and Experimental Physics, Moscow} 
  \author{T.~Aziz}\affiliation{Tata Institute of Fundamental Research, Mumbai} 
  \author{S.~Bahinipati}\affiliation{University of Cincinnati, Cincinnati, Ohio 45221} 
  \author{A.~M.~Bakich}\affiliation{University of Sydney, Sydney, New South Wales} 
% ...
\collaboration{The Belle Collaboration}

\maketitle

\tighten

{\renewcommand{\thefootnote}{\fnsymbol{footnote}}}
\setcounter{footnote}{0}

In the standard model (SM) framework, $CP$ violation arises 
only from the irreducible phase in the Cabibbo-Kobayashi-Maskawa (CKM) matrix~\cite{ckm}.
$CP$ violation in flavor changing neutral currents (FCNC) is
sensitive to new physics (NP), as new particles beyond the SM may be involved in loop diagrams.
Such additional loop diagrams potentially add new $CP$ phases and induce deviations from
the SM expectation for time-dependent $CP$ asymmetries~\cite{bib:lucy}.

In the decay chain $\Upsilon(4S)\to\bbbar$,
one of the two $B$ mesons decays into a $CP$ eigenstate $f_{CP}$
at time $t_{CP}$, and the other decays into a flavor specific
state $f_{\rm tag}$ at time $t_{\rm tag}$.
The $CP$ violation parameters are measured using the
time-dependent decay rate~\cite{sanda}
\begin{eqnarray}
{\cal P}(\Delta t)
& = & \frac{e^{-|\Delta t|/\tau_{B^0}}}{4\tau_{B^0}}
[1 + q \cdot \{ {\cal S}_{f} \sin(\Delta m_d \Delta t) \nonumber \\
 & &         + {\cal A}_{f} \cos(\Delta m_d \Delta t ) \} ], 
\label{eq:signal-pdf}
\end{eqnarray}
where $\dt=t_{CP} - t_{\rm tag}$,
$\tau_{B^0}$ is the $B^0$ lifetime, $\dmd$ is 
the $B^0\overline{B}{}^0$ mixing frequency
and $q=+1$ $(-1)$ when $f_{\rm tag} = B^0(\overline{B}{}^0)$.
The parameters
${\cal S}_{f}$ and ${\cal A}_{f}$ represent
mixing-induced and direct $CP$-violation, respectively,

In the SM,
the $CP$ violation parameters in $b\to s$ ``penguin" 
and $b\to c$ ``tree" transitions
are predicted to be ${\cal S}_{f}\simeq -\xi_{f}\sin 2\phi_1$
and ${\cal A}_{f}\simeq 0$
with small theoretical uncertainties~\cite{b2s},
where $\phi_1$ is one of the CKM weak angles
and $\xi_{f}$ is the $CP$ eigenvalue of the final state.
Recent measurements~\cite{belle-b2s06,babar-b2s,kspizpiz-babar}
however, indicate that the effective $\sin 2\phi_1$ value,
$\sin 2\phi_1^{\rm eff}$, measured with penguin processes
is different from $\sin 2\phi_1=0.687\pm0.025$ measured in tree
decays by $2.6$ standard deviations ($\sigma$)~\cite{hfag}.
New particles in the loop diagrams may have shifted the weak phase.

Belle and BaBar recently observed $b\to d$ FCNC transitions in $B$ meson decays into
two kaons $B^+\to K^{+}\ks$ and $B^0\to\ksks$~\cite{b2kk-belle, b2kk-babar, CC}.
The $CP$ violation parameters in $b\to d$ transitions
are expected to be small because the weak phases in $\bbbar$ mixing and the $b\to d$
transition  cancel~\cite{ksks-np}.
Thus a measurement of large $CP$ violation could indicate 
the existence of NP.
It has been also pointed out that the $CP$ violation parameters
as well as the branching fraction for $B^0\to\ksks$ decay can be used to
constrain $\phi_2$ ~\cite{phi2-ksks}.

In this report, we present measurements of $CP$ violation parameters in $B^0\to\kspizpiz$ decays
that proceed through $b\to s \overline{q}q$ ($q=u,d$) transitions,
and $B^0\to\ksks$ decays in which pure $b\to d\overline{s}s$
penguin processes are the dominant contribution.
Because the $\xi_{f}$ for $B^0\to\kspizpiz$ decays is $+1$ 
for any intermediate resonance~\cite{Gershon:2004tk}, 
the SM expectations for the $CP$ violation parameters are ${\cal A}\sim0$ 
and ${\cal S}=-\sin 2\phi_1$ with small theoretical uncertainty~\cite{Cheng-dads}.
Both analyses employ a technique~\cite{ks-vertex} that
reconstructs the $B^0$ decay vertex position using only 
$\ks$'s decaying into $\pi^+\pi^-$.
The $CP$ violation parameters of these $B$ decay modes have been 
measured by BaBar~\cite{kspizpiz-babar, b2kk-babar}
with 227 (348)$\times 10^6$ $B\overline{B}$ pairs
for $B^0\to \kspizpiz$ ($\ksks$).
They found a $2.2\sigma$ deviation from the SM expectation
for $B^0\to\kspizpiz$ and a large ${\cal S}_{\ksks}$ value;
however, no clear conclusion about NP can be obtained because of the large statistical errors.

We use a data sample containing $657 \times 10^6 B\overline{B}$ pairs
collected by the Belle detector~\cite{belle-detector}
at the KEKB $e^+e^-$ asymmetric-energy (3.5 on 8 GeV) collider
operating at $\Upsilon(4S)$ resonance~\cite{kekb}.
The $\Upsilon(4S)$ is produced with a Lorentz boost factor
of $\beta\gamma = 0.425$ approximately along the electron beamline ($z$).

Since the two $B$ mesons are produced nearly at rest
in the $\Upsilon(4S)$ center-of-mass system (cms),
we determine $\dt = \Delta z / \beta\gamma c$,
where $\Delta z$ is the distance between the two $B$ decay positions
in the $z$ direction.
In the Belle detector,
a silicon vertex detector (SVD) and a 50-layer central drift chamber (CDC)
are used for charged particle tracking.
Photons are detected with an electromagnetic calorimeter (ECL) 
comprised of CsI(Tl) crystals.
The devices are placed inside a superconducting solenoid coil
providing a 1.5~T magnetic field.

We reconstruct $\ks \to \pi^+\pi^-$ candidates
from pairs of oppositely charged tracks.
The candidates are required to satisfy the following conditions:
(a) the two charged tracks do not originate from
the beam interaction point (IP),
(b) the $z$ positions of the two charged tracks at
the $\ks$ decay position have to match,
(c) the direction of the pion pair momentum is consistent with
the direction from the IP to the $\ks$ decay position, and 
(d) the invariant mass of the two charged tracks 
is within $\pm15$~MeV/c$^2$ of the nominal $\ks$ mass.
The criteria (a) to (c) are optimized
for each $B$ decay mode, and depend on $\ks$ momentum.
We form $\pi^0\to\gamma\gamma$ candidates from pairs of two photons.
The photon energy in the laboratory frame is required to be greater
than 50~MeV. The $\pi^0$ candidates should have invariant masses
in the range $0.118$~GeV/c$^2 < M_{\gamma\gamma} < 0.150$~GeV/c$^2$.

We reconstruct a $B^0$ candidate by combining a $\ks\to\pi^+\pi^-$ and 
two $\pi^0$s (a $\ks\to\pi^+\pi^-$) 
for $B^0\to\kspizpiz$ ($\ksks$) decay.
Two kinematic variables are used for the $B^0$ candidate selection:
$\de = E_{B} - E_{\rm beam}$ 
and $\mbc=\sqrt{E_{\rm beam}^2 - p_B^2}$,
where $E_{\rm beam}$ is the cms beam energy, and
$E_{B}$ and $p_{B}$ are the cms energy and momentum, 
respectively, of a $B$ candidate.
We require $|\de|<0.25$ ($0.20$)~GeV and $\mbc>5.20$~GeV/c$^2$
for $B^0\to\kspizpiz$ ($\ksks$).

We find 34\% ($<1$\%) of events have multiple $B^0\to\kspizpiz$ ($\ksks$) candidates.
We choose the candidate having the smallest $\chi^2$ value for the $\pi^0$
for $B^0\to\kspizpiz$, and
having the smallest difference of $M_{\pi^+\pi^-}$ from the nominal $\ks$ mass 
for $B^0\to\ksks$.
For $B^0\to\kspizpiz$ candidates, we find 23\% of signal events
in the Monte-Carlo (MC) simulation are incorrectly reconstructed
self-cross-feed (SCF) events;
for events at least one particle is replaced
with one from either the accompanying $B$ meson decay 
or from beam background.
For 99.5\% of the SCF events, only photons are replaced;
the $B$ vertex position is always reconstructed with the correct $\ks$.
Therefore, we do not distinguish between the SCF and correctly-reconstructed
$B$ candidates for this analysis.
No SCF component is found for $B^0\to\ksks$ decays.

We remove $B^0\to\kspizpiz$ candidates satisfying one of
the following conditions: 
$M_{\ks\pi^0}>4.8$~GeV/c$^2$ to suppress $B^0\to\ks\pi^0$ decays,
$1.77$~GeV/c$^2 < M_{\ks\pi^0} < 1.94$~GeV/c$^2$ ($D^0$ veto), 
$3.27$~GeV/c$^2 < M_{\pi^0\pi^0} < 3.49$~GeV/c$^2$ ($\chi_{c0}$ veto)
and $\cos\theta_{\gamma}>0.9$,
where $\theta_{\gamma}$ is the angle between the $\pi^0$ boost direction from
the laboratory frame and one of the photons in the $\pi^0$ rest frame.
We impose the last condition to suppress $b\to s\gamma$ backgrounds.

The dominant background is the continuum, $e^+e^-\to q\overline{q}$
($q=u,d,s,c$).
To suppress this background, 
we form a likelihood ratio 
${\cal R} = L_{\rm sig} / (L_{\rm sig} + L_{\rm bg})$,
where $L_{\rm sig (bg)}$ is a likelihood function for
the signal (continuum) based on the event topology
and the $B^0$ flight direction with respect to the beam axis
in the cms.
The $B^0\to\kspizpiz$ ($\ksks$) candidates having
${\cal R}>0.1 (0.25)$ are selected.
We also use ${\cal R}$ to discriminate
the signal and background in the fits.

We identify the flavor of $f_{\rm tag}$ using an algorithm~\cite{flavor-tag} 
that provides two variables: $q$ defined in Eq.~(\ref{eq:signal-pdf})
and $r$.
The parameter $r$ ranges from $r=0$ for no flavor discrimination
to $r=1$ for an unambiguous flavor determination. 
Candidate events having $r>0.1$ are
divided into six $r$-bins ($\ell=1,6$).
The wrong tag fraction in each $\ell$ bin, $w_{\ell}$,
and their differences between $B^0$ and $\overline{B}{}^0$, $\Delta w_{\ell}$,
are determined using data~\cite{sss06-belle}.

We apply the vertex reconstruction algorithm of Ref.~\cite{ks-vertex}.
The vertex position of the $f_{CP}$ decay is determined
using the $\ks\to\pi^+\pi^-$ pseudo-track and the IP profile.
The two charged tracks are required to have enough associated SVD hits
for vertex reconstruction.
We reconstruct the $B^0\to f_{\rm tag}$ decay vertex position
using charged tracks that are not used for the $f_{CP}$ reconstruction.
We measure both ${\cal S}_{f}$ and ${\cal A}_{f}$ parameters
with candidate events in which the vertex position is reconstructed
successfully.
Candidates with no vertex position information
are used only for ${\cal A}_{f}$ measurements.
Using a signal MC sample, 
we estimate a vertex reconstruction efficiency of
45\% (56\%) for $B^0\to\kspizpiz$ ($\ksks$) decays.

We determine signal yields with an extended unbinned maximum likelihood
fit that makes use of the variables $\de$, $\mbc$ and ${\cal R}$.
The selected candidate events include not only the signal 
but also continuum and $B\overline{B}$ backgrounds.

We use a two-dimensional smoothed $\de$-$\mbc$ histogram
for the $B^0\to\kspizpiz$ signal probability density function (PDF),
while we model the $\de$ ($\mbc$) shape using
a sum of two Gaussians (a single Gaussian) for $B^0\to\ksks$ decay.
Histogram PDFs are employed for the ${\cal R}$ distributions
of signal in both decay modes.
The PDFs are determined using signal MC simulations.

The $\de$ ($\mbc$) shape of the continuum background
is modeled as a polynomial (an ARGUS~\cite{argus}) function.
The parameters of the background functions are floated in the fit.
An ${\cal R}$ PDF for continuum background is determined from a data sample
collected at a center of mass energy 
60~MeV below the $\Upsilon(4S)$ resonance (off-resonance)
for the $B^0\to\kspizpiz$
and events in the sideband region
($\mbc<5.26$~GeV/c$^2$) for $B^0\to\ksks$ decay. 
We use the off-resonance data for the former decay mode, since
$B\overline{B}$ background contaminates the sideband
region, while the latter decay mode has a negligibly small 
contribution from $B\overline{B}$ background.
A large MC sample of $B\overline{B}$ background is used
to obtain a smoothed histogram $\de$-$\mbc$ PDF
and a histogram PDF of ${\cal R}$ for 
$B^0\to\kspizpiz$ decay.

\begin{figure}
\begin{center}
\includegraphics[width=0.3\textwidth]{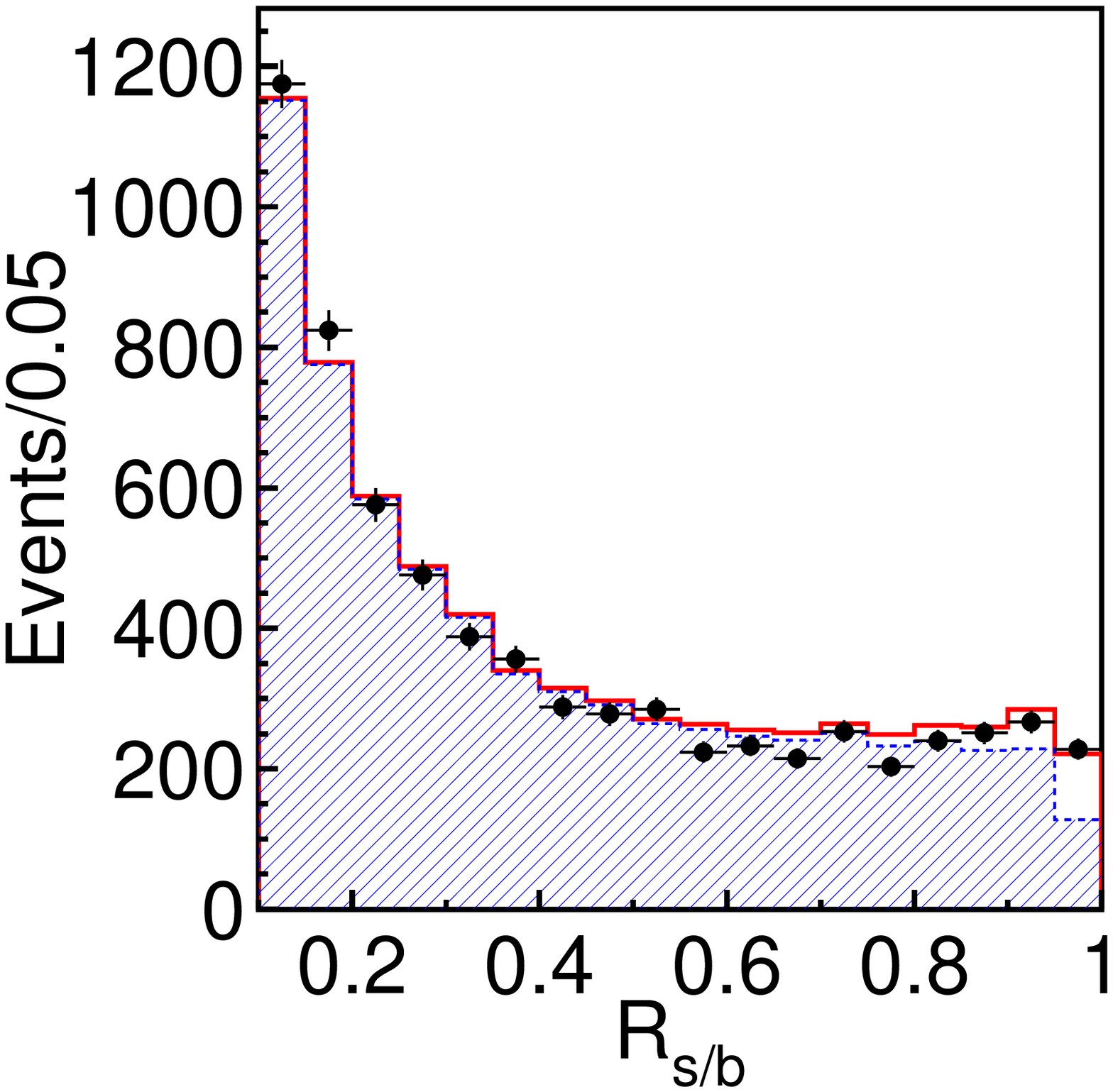}
\includegraphics[width=0.3\textwidth]{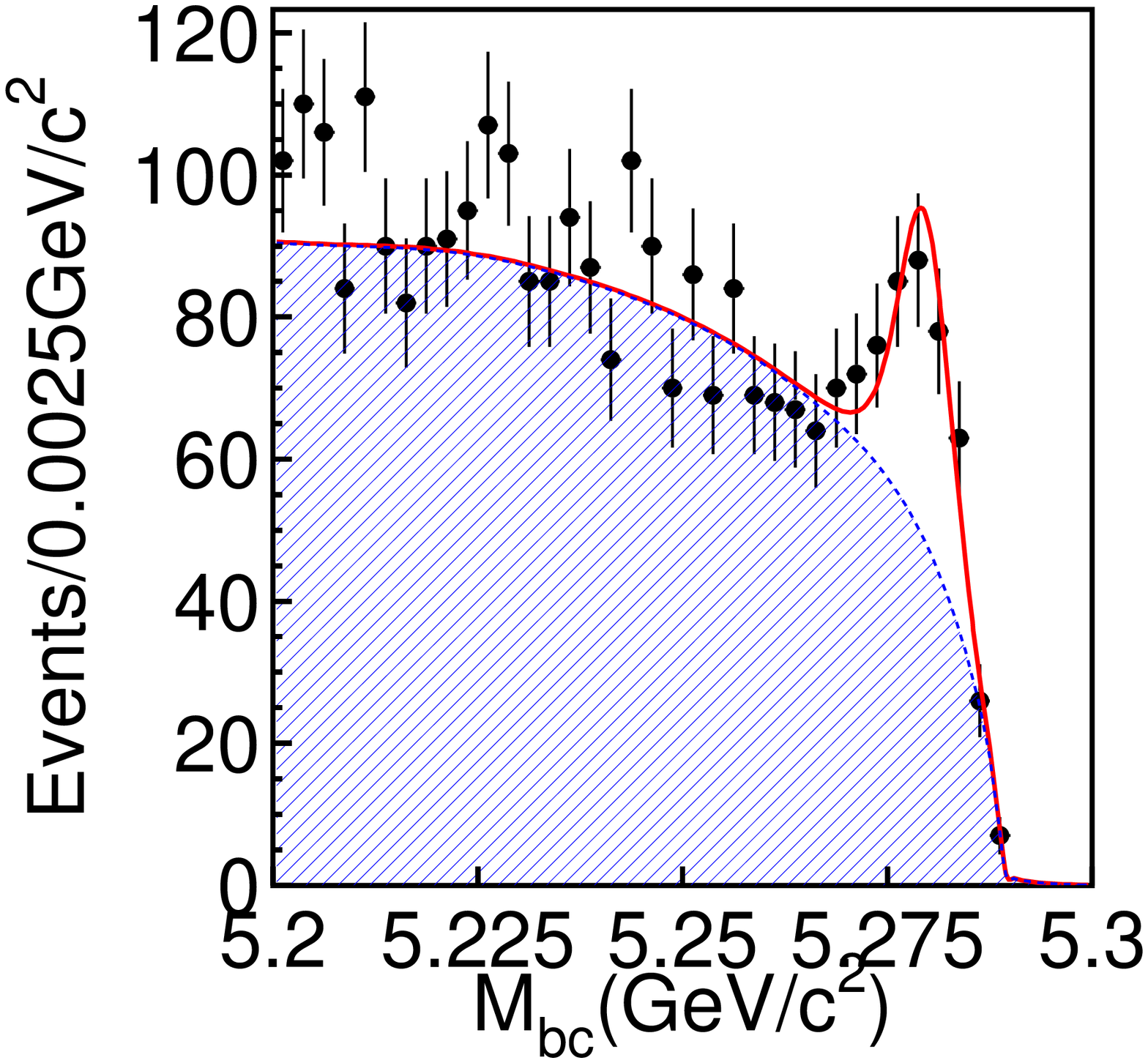}
\includegraphics[width=0.3\textwidth]{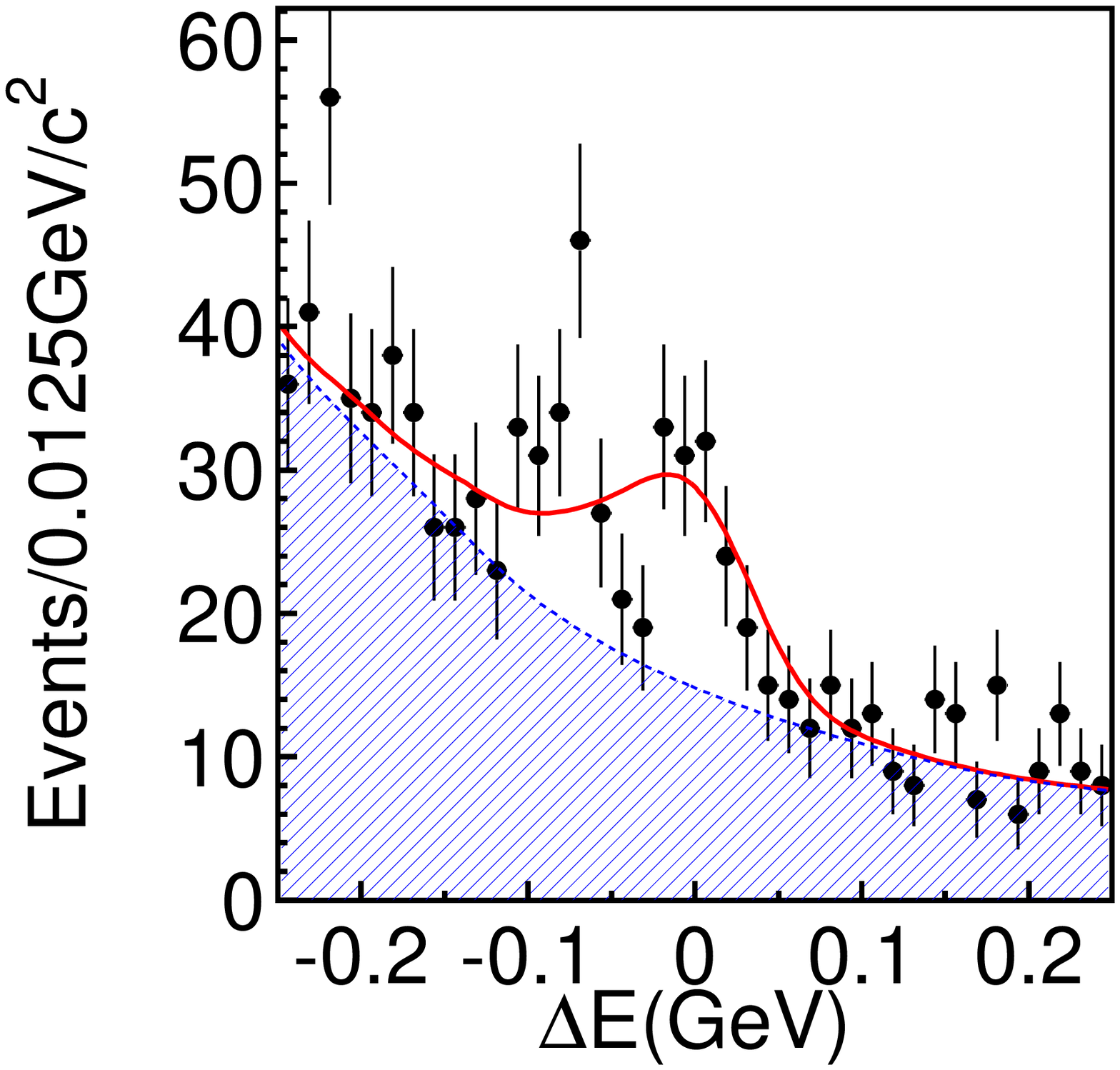}
\includegraphics[width=0.3\textwidth]{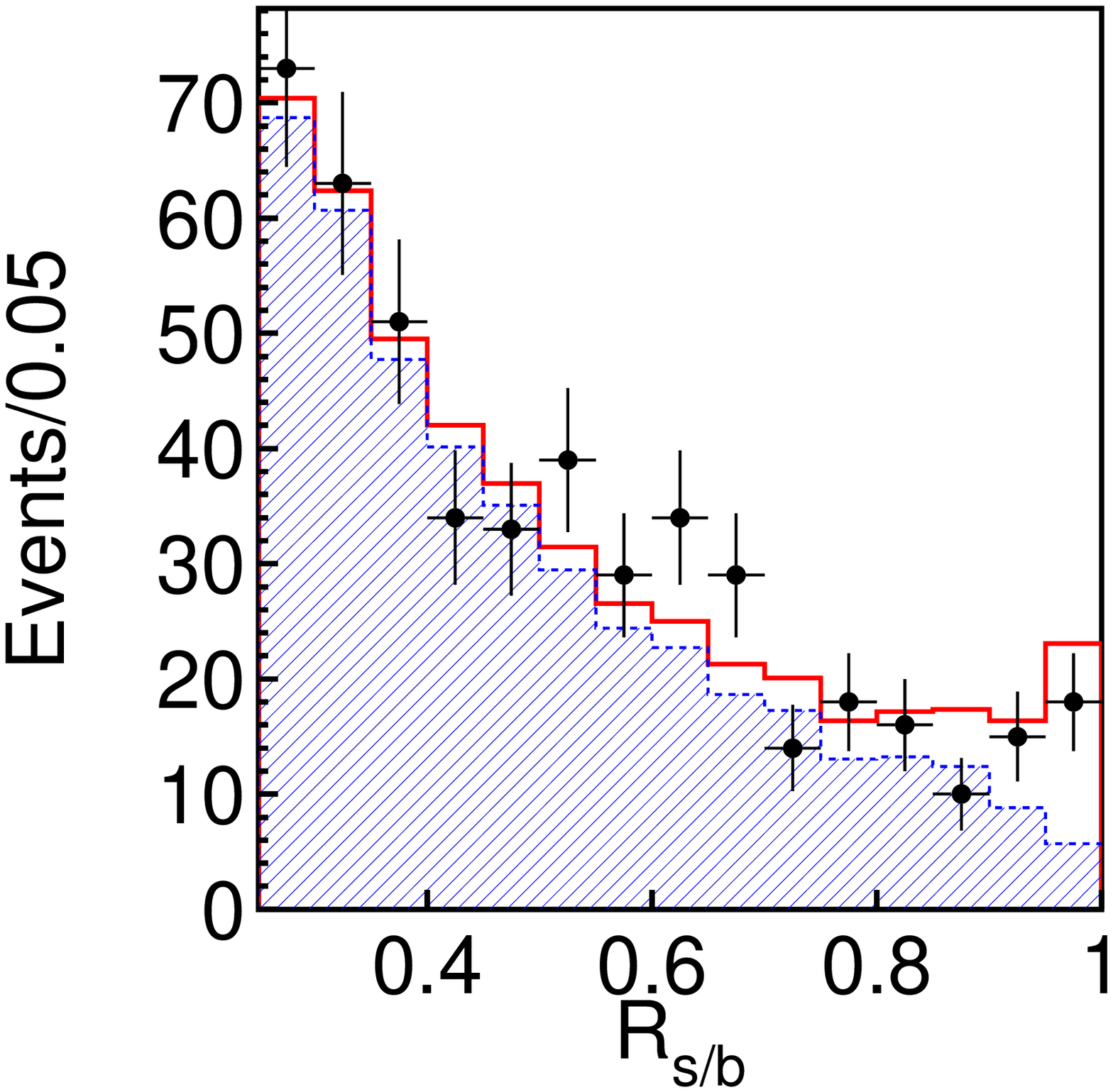}
\includegraphics[width=0.3\textwidth]{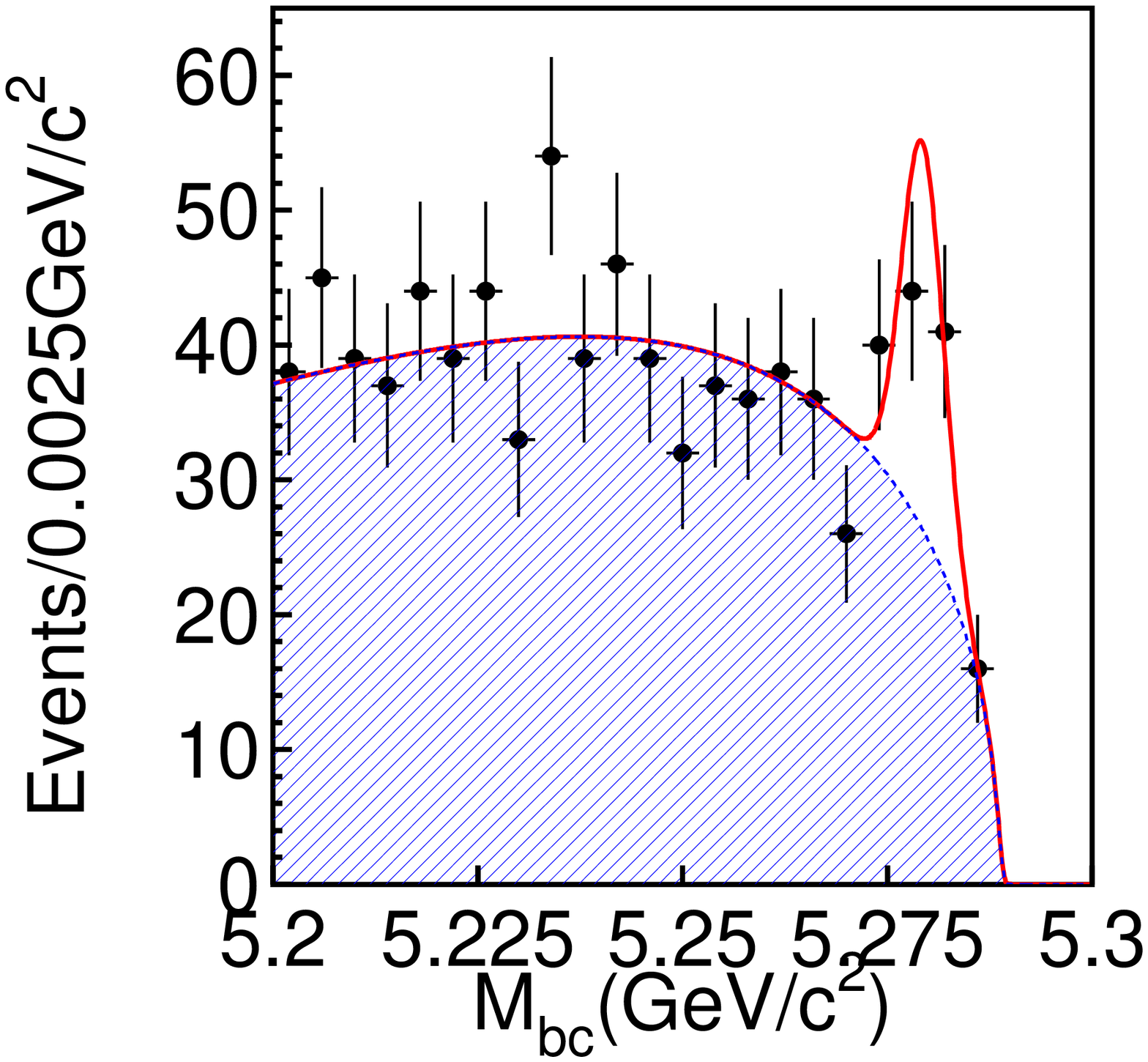}
\includegraphics[width=0.3\textwidth]{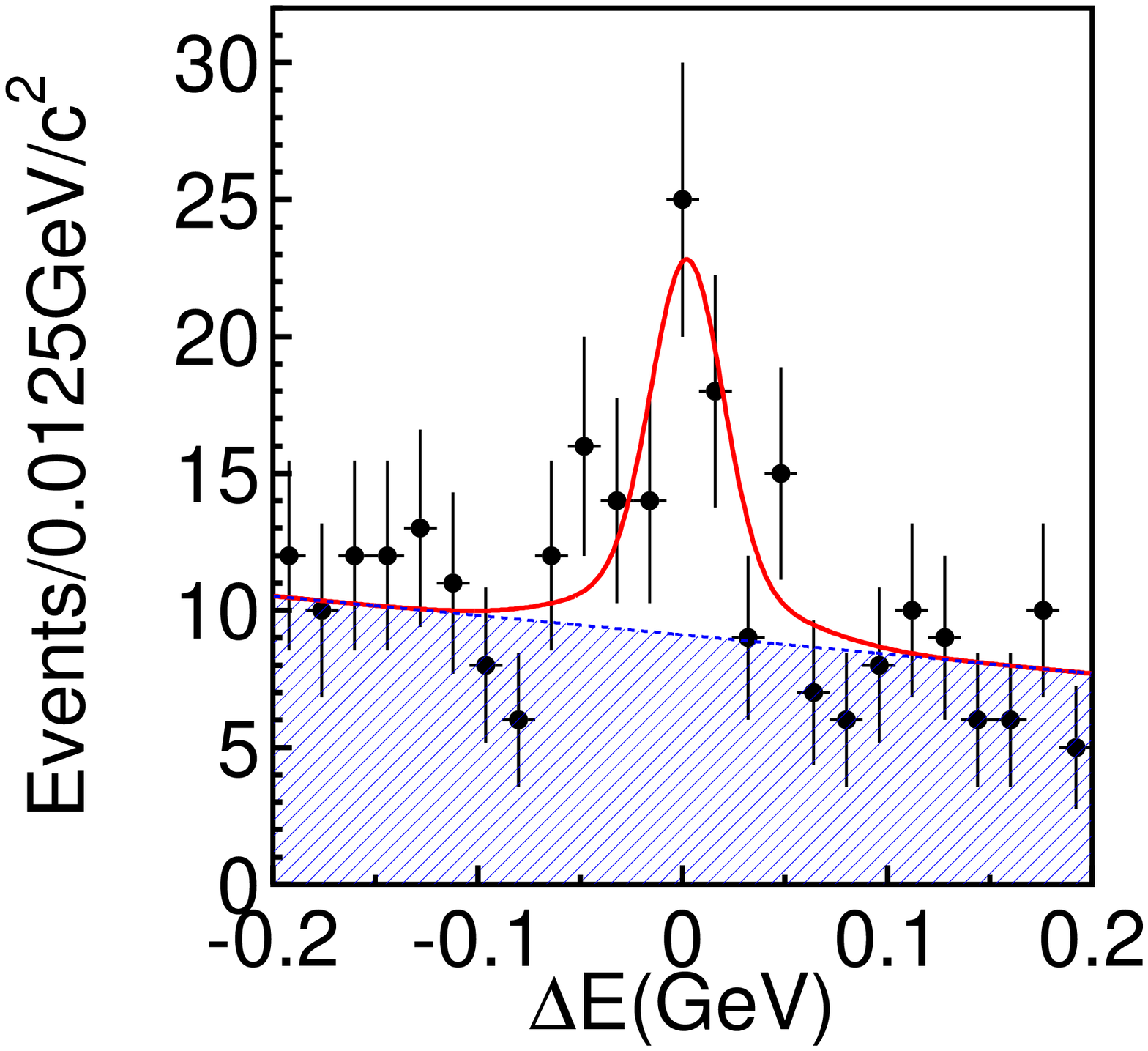}
\rput(-11.1,8){(a)}
\rput(-6.0,8){(b)}
\rput(-0.9,8){(c)}
\rput(-11.1,3.7){(d)}
\rput(-6.0,3.7){(e)}
\rput(-0.9,3.7){(f)}
\caption{
(a) distributions of ${\cal R}$ for the $\de$ and $\mbc$ signal region,
(b) $\mbc$ for the $\de$ signal region with ${\cal R} > 0.9$, and
(c) $\de$ for the $\mbc$ signal region with ${\cal R} > 0.9$
for $B^0\to\ks\pi^0\pi^0$.
(d) distributions of ${\cal R}$ for the $\de$ and $\mbc$ signal region,
(e) $\mbc$ for the $\de$ signal region with ${\cal R} > 0.6$,
(f) $\de$ for the $\mbc$ signal region with ${\cal R} > 0.6$
for $B^0\to\ks\ks$.
The solid curves and histograms show the fits to signal plus background distributions,
and hatched areas show the background contributions.
}
\label{fig:de-mbc-r}
\end{center}
\end{figure}

By fitting the data, we determine the signal yields in the signal box.
The signal box is defined as $-0.15$~GeV$ < \de < 0.1$~GeV
and $\mbc>5.27$~GeV/c$^2$ ($|\de|<0.1$~GeV and $\mbc>5.27$~GeV/c$^2$)
for $B^0\to\kspizpiz$ ($\ksks$) decay.
For $B^0\to\kspizpiz$ decay,
we divide the candidate events into two categories:
events with and without vertex information, and
fit the data subsets separately to take into account
a possible difference of the signal fraction.
For $B^0\to\ksks$ decay, instead of separating the candidate events
in the fit, 
we perform a fit to the whole data sample and
employ the vertex reconstruction efficiency
estimated from a MC simulation (sideband events)
for the signal (continuum) events
to determine the signal fractions for the two categories.
We obtain $129\pm21$ ($178\pm24$) $B^0\to\kspizpiz$ events
with (without) the vertex information,
which is consistent with the expected vertex reconstruction efficiency,
and $58\pm11$ $B^0\to\ksks$ events.
Figure~\ref{fig:de-mbc-r} shows the projections of $\de$, $\mbc$ and ${\cal R}$
for candidate events.

We determine the $CP$ violation parameters
from an unbinned maximum likelihood fit to the $\dt$ distribution of
the candidate events in the signal box.
For each event, we calculate a likelihood value:
\begin{eqnarray}
P_i & =& (1-f_{\rm ol})\sum_k f_k(\de,\mbc,{\cal R})
      {\cal P}_k(\dt_i) \otimes R_k(\dt_i) \nonumber \\
    & & + f_{\rm ol}{\cal P}_{\rm ol}(\dt_i),
\label{eq:likelihood}
\end{eqnarray}
where $k$ indicates signal, continuum ($\qq$) and $B\overline{B}$ background
components, $f_k(\de,\mbc,{\cal R})$ is the fraction of the component $k$
as a function of $\de, \mbc, {\cal R}$.
For $B^0\to\ksks$, the $B\overline{B}$ contribution is neglected, i.e.
$f_{B\overline{B}}=0$.
For the signal function ${\cal P}_{\rm sig}(\dt)$,
we use Eq.~(\ref{eq:signal-pdf})
modified to incorporate the flavor mis-assignment effect
using $w_{\ell}$ and $\Delta w_{\ell}$.
This distribution is convolved with a resolution function 
$R_{\rm sig}(\dt)$~\cite{res-func} to obtain the signal PDF.
The continuum $\dt$ distribution ${\cal P}_{\qq}(\dt)$ 
consists of prompt and lifetime components, which is convolved
with a resolution function $R_{\qq}(\dt)$ comprised of
two Gaussians;
all the parameters of the functions are determined
using the off-resonance (sideband) data 
for $B^0\to\kspizpiz$ ($\ksks$) decay.
We employ a large MC sample to obtain ${\cal P}_{B\overline{B}}(\dt)$
and $R_{B\overline{B}}(\dt)$; the same functional forms as those of the continuum background
 are used.
All the PDFs are combined with a outlier PDF ${\cal P}_{\rm ol}(\dt)$
that takes into account a small fraction ($f_{\rm ol}$) 
of events having large $\dt$ values, giving a likelihood value $P_i$
for $i$-th candidate event. 
For candidate events with no vertex information,
we use the likelihood function given in Eq.~(\ref{eq:likelihood}) integrated over $\dt$.

\begin{figure}
\begin{center}
\includegraphics[width=0.45\textwidth]{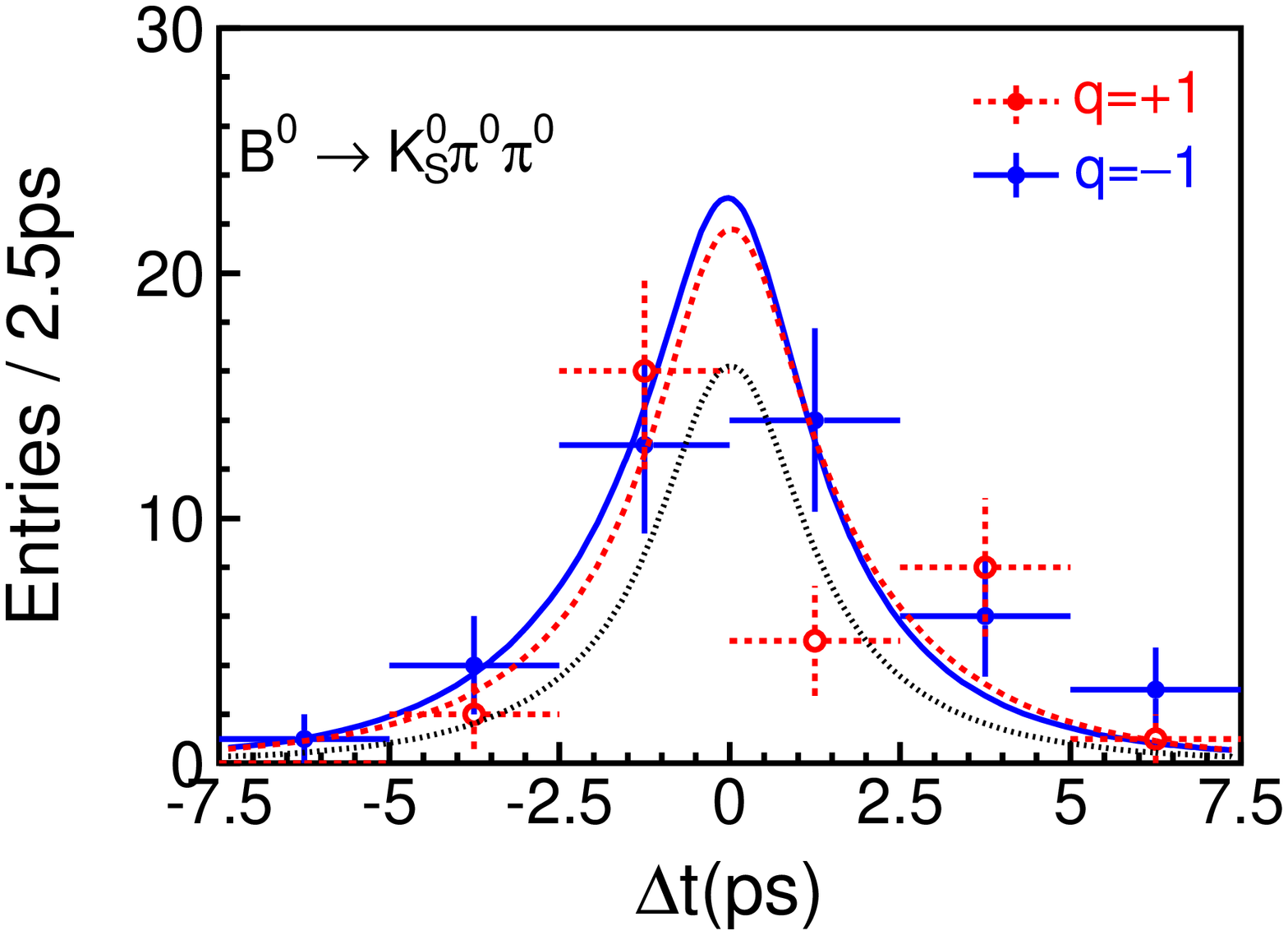} 
\includegraphics[width=0.45\textwidth]{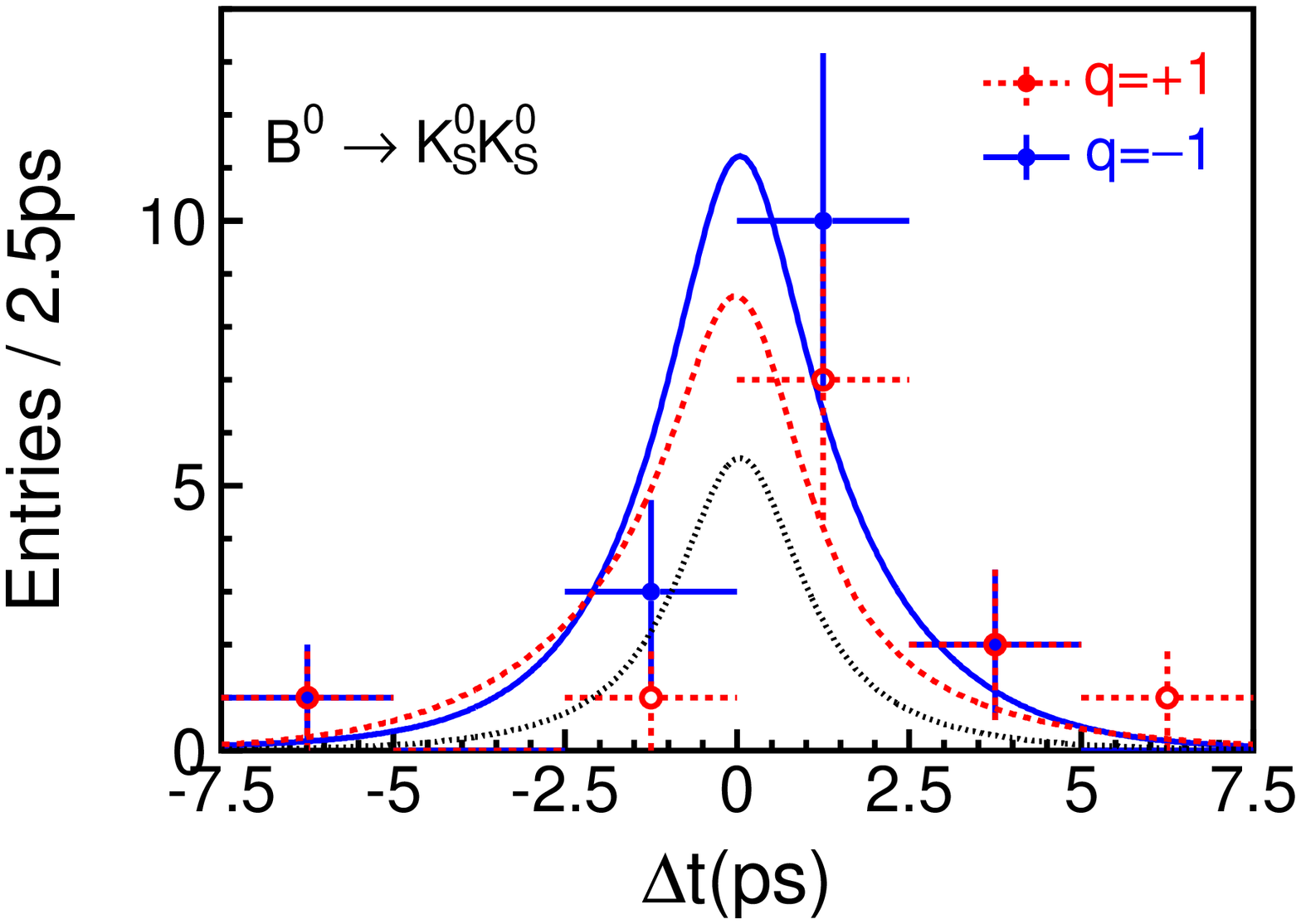}
\includegraphics[width=0.45\textwidth]{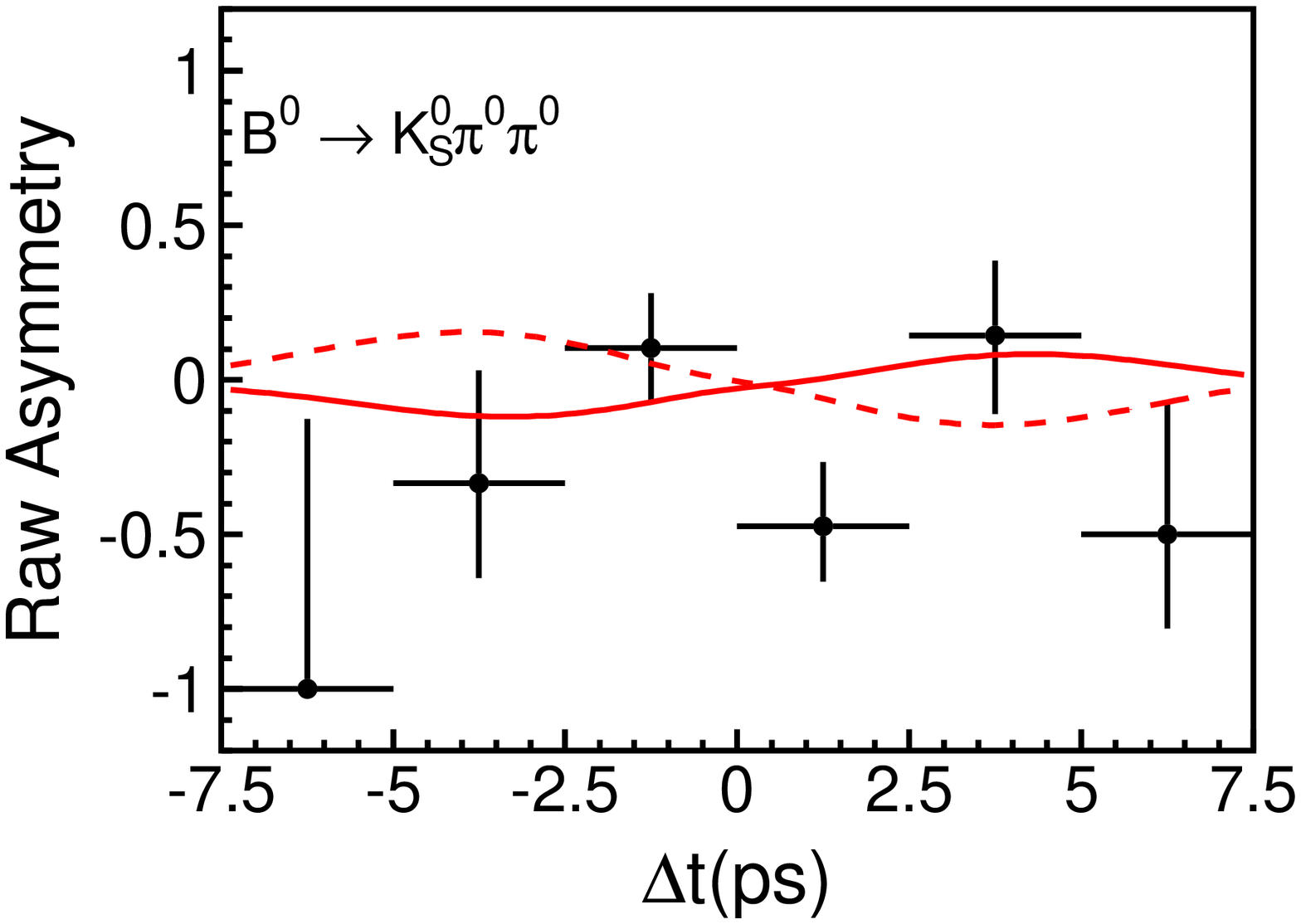} 
\includegraphics[width=0.45\textwidth]{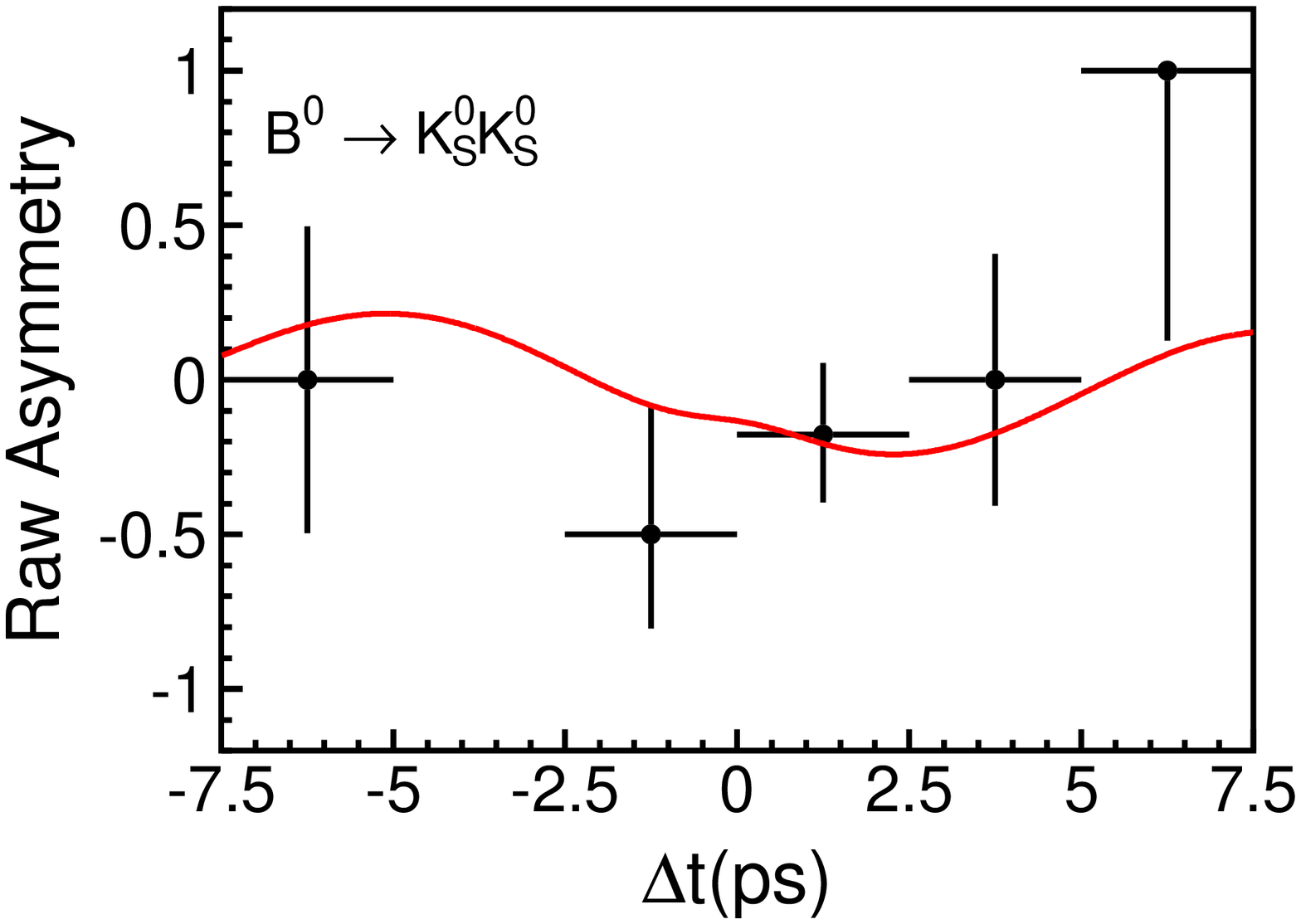}
\rput(-13.4,10.2){(a)}
\rput(-5.9,10.2){(b)}
\rput(-13.4,4.6){(c)}
\rput(-5.9,4.6){(d)}
\caption{
$\dt$ distributions and asymmetries for events with good tags ($r>0.5$)
for (a),(c) $B^0\to\ks\pi^0\pi^0$ with ${\cal R} >0.9$ and (b), (d) $B^0\to\ks\ks$ 
with ${\cal R} >0.6$.
In the $\dt$ plots, the dashed and solid lines show the fit result for events with $q=\pm1$.
The dotted lines show the background contribution.
In the asymmetry plots, solid curves show the fit results;
the dashed curve in (c) shows the SM expectation ${\cal A}=0$ and ${\cal S}=-\sin 2\phi_1$.
}
\label{fig:dt-acp}
\end{center}
\end{figure}

In the fit, ${\cal S}_{f}$ and ${\cal A}_{f}$
are the only free parameters and are determined
by maximizing ${\cal L} = \prod_i P_i$.
The fit yields 
$\skspizpiz=\skspizpizvalue \pm \skspizpizstat$,
$\akspizpiz=\akspizpizvalue \pm \akspizpizstat$,
$\sksks=\sksksvalue \pm \sksksstat$ and
$\aksks=\aksksvalue \pm \aksksstat$,
where the errors are statistical.
Figure~\ref{fig:dt-acp} show the $\dt$ distributions
and the asymmetry ${\cal A}_{CP}$ in each $\dt$ bin,
where ${\cal A}_{CP}= (N_+ - N_-)/(N_+ + N_-)$
and $N_{+(-)}$ is the number of candidate events with
$q=+1$ ($-1$).

\begin{table}
\begin{center}
\caption{Systematic errors of the measured $CP$ violation parameters.}
\begin{tabular}{ccccc}
\hline
\hline

sources & $\skspizpiz$ & $\akspizpiz$ & $\sksks$ & $\aksks$ \\
\hline
wrong tag fraction & $\pm 0.01$ & $\pm 0.01$ & $\pm 0.02$ & $\pm 0.01$ \\ 
physics parameters  & $<0.01$ & $<0.01$ & $\pm 0.01$ & $\pm 0.01$ \\ 
resolution function & $\pm 0.04$ & $\pm 0.02$ & $\pm 0.06$ & $<0.01$ \\ 
background $\dt$ shape & $\pm 0.05$ & $\pm 0.02$ & $\pm 0.04$ & $\pm 0.02$ \\ 
background fraction & $\pm 0.05$ & $\pm 0.03$ & $\pm 0.04$ & $\pm 0.02$ \\ 
possible fit bias & $\pm 0.02$ & $\pm 0.01$ & $\pm 0.02$ & $\pm 0.01$ \\ 
vertex reconstruction & $\pm 0.01$ & $\pm 0.02$ & $\pm 0.01$ & $\pm 0.02$ \\ 
tag side interference & $<0.01$ & $\pm 0.04$ & $<0.01$ & $\pm 0.03$ \\ 
\hline
total & $\pm 0.09$ & $\pm 0.06$ & $\pm 0.08$ & $\pm 0.05$ \\ 
\hline
\hline
\label{table:systematic}
\end{tabular}
\end{center}
\end{table}

Table~\ref{table:systematic} shows the systematic errors
for the measured $CP$ violation parameters.
For the uncertainties on the wrong tag fractions,
we vary the wrong tag fraction parameters, 
$w_{\ell}$ and $\Delta w_{\ell}$, by $\pm 1\sigma$ individually,
and sum up the $CP$ violation parameter variations in quadrature.

The parameters $\taub$ and $\dmd$ are varied by
their errors~\cite{pdg} to determine the
systematic uncertainties from the physics parameters.

We perform fits by varying each parameter in the signal resolution function 
$R_{\rm sig}(\dt)$.
The differences between the fitted parameters and the nominal values
are added in quadrature.

We vary the background $\dt$ PDF parameter values by $\pm 1\sigma$,
and take the quadratic sum of the $CP$ violation parameter variations
as the systematic error.
For $B^0\to\kspizpiz$, we also take into account the uncertainties originating 
from possible $CP$ violation in the $B\overline{B}$ background.
Using a large MC sample, we find 18\% of the charmless $B$ decay background
are $CP$ eigenstates.
Each $CP$ asymmetry parameter is varied between $\pm 1$, and the differences
are added in quadrature for the systematic error. 

To estimate the background fraction systematic error,
we vary each parameter in the PDFs of $\de$, $\mbc$ and ${\cal R}$,
and add the differences in quadrature.
For $B^0\to\kspizpiz$,
we also take into account uncertainties in the $B\overline{B}$
background yields;
we vary the charm (charmless) $B$ decay background 
yields by $\pm 2\sigma$ ($\pm 100$\%).
For $B^0\to\ks\pi^0\pi^0$, systematic uncertainties coming from the
possible signal shape difference due to the uncertainty in the intermediate resonant state
are also included.
We generate pseudo-experiment MC samples based on PDFs assuming 100\% resonant decays
for the cases of $B^0\to f_0\ks$ and $B^0\to K^{*0}\pi^0$ decays,
fit the samples with the nominal PDFs, and take the differences
as systematic errors.

A possible fit bias is examined by fitting a large number of MC events.
We find no statistically significant bias and assign the MC statistical error as the
systematic error.

The systematic uncertainties for the vertex reconstruction
are estimated by changing the charged track selection criteria,
criteria on the vertex fit $\chi^2$ requirements,
small bias corrections on $\Delta z$, smearing due to the $B$ flight length,
and the SVD mis-alignment parameters.

Finally, we take into account the possible $CP$ violation effect
in $B^0\to f_{\rm tag}$ decay~\cite{tsi}.

To validate our measurements,
various checks are carried out.
We measure the $B^0$ lifetime;
we obtain $\taub = 1.32\pm0.27$ ($1.58\pm0.44$)~ps
for $B^0\to\kspizpiz$ ($\ksks$) decay,
in agreement with the world average (WA) value~\cite{pdg}.
We perform a fit to sideband events; no asymmetries are found for both decay modes.
We measure the branching fractions
of $B^0\to D^0(\to K_S^0 \pi^0)\pi^0$ 
and $B^0\to\ksks$; we obtain $(3.3\pm0.4)\times 10^{-4}$
and $(1.1\pm0.2)\times 10^{-6}$, respectively, consistent with the WA values~\cite{pdg, hfag}.
The errors are statistical only.

In summary we measure the $CP$ violation parameters
in FCNC $B^0$ decays into $\kspizpiz$ and $\ksks$ 
using a data sample containing $657\times 10^6$ $B\overline{B}$ pairs.
We obtain
\begin{eqnarray*}
\skspizpiz& =& \skspizpizvalue \pm \skspizpizstat \pm \skspizpizsyst, \\
\akspizpiz& =& \akspizpizvalue \pm \akspizpizstat \pm \akspizpizsyst, \\
\sksks& =& \sksksvalue \pm \sksksstat \pm \skskssyst~~ {\rm and}, \\
\aksks& =& \aksksvalue \pm \aksksstat \pm \akskssyst,
\end{eqnarray*}
where the first and second errors are statistical and systematic,
respectively.
We estimate that the deviation 
from the SM expectation for $CP$ violation parameters in $B^0\to\kspizpiz$ decay
${\cal A}=0$ and ${\cal S}=\sin 2\phi_1$ has a significance of $2.0\sigma$.
No $CP$ asymmetry is found in $B^0\to\ksks$ decay,
which is consistent with the SM prediction.
Our results are consistent with other
measurements~\cite{kspizpiz-babar, b2kk-belle, b2kk-babar}.

We thank the KEKB group for the excellent operation of the
accelerator, the KEK cryogenics group for the efficient
operation of the solenoid, and the KEK computer group and
the National Institute of Informatics for valuable computing
and Super-SINET network support. We acknowledge support from
the Ministry of Education, Culture, Sports, Science, and
Technology of Japan and the Japan Society for the Promotion
of Science; the Australian Research Council and the
Australian Department of Education, Science and Training;
the National Science Foundation of China and the Knowledge
Innovation Program of the Chinese Academy of Sciences under
contract No.~10575109 and IHEP-U-503; the Department of
Science and Technology of India; 
the BK21 program of the Ministry of Education of Korea, 
the CHEP SRC program and Basic Research program 
(grant No.~R01-2005-000-10089-0) of the Korea Science and
Engineering Foundation, and the Pure Basic Research Group 
program of the Korea Research Foundation; 
the Polish State Committee for Scientific Research; 
the Ministry of Education and Science of the Russian
Federation and the Russian Federal Agency for Atomic Energy;
the Slovenian Research Agency;  the Swiss
National Science Foundation; the National Science Council
and the Ministry of Education of Taiwan; and the U.S.\
Department of Energy.


\begin{thebibliography}{99}
\bibitem{CC}
Throughout this paper, 
the inclusion of the charge conjugate mode decay is implied
unless otherwise stated.
\end{thebibliography}
\end{document}